\documentclass{elsart3}
\usepackage{natbib}
\usepackage{graphicx}

\newcommand {\mm}       {\ensuremath}%
\newcommand {\mr}       {\mathrm}%
\newcommand {\vu}    [2] {\mm{#1\:\mr{#2}}}                 
\newcommand {\GeV}  {\ifmmode{\mr{Ge\kern -0.1em V}}%
                    \else\textrm{Ge\kern -0.1em V}\fi}%
\newcommand {\TeV}  {\ifmmode{\mr{Te\kern -0.1em V}}%
                    \else\textrm{Te\kern -0.1em V}\fi}%
\newcommand{\bc}{\begin{center}}
\newcommand{\ec}{\end{center}}
\newcommand{\bi}{\begin{itemize}}
\newcommand{\ei}{\end{itemize}}

\newcommand{\volp}{\vspace{7mm}}
\newcommand{\vups}{\vspace{6mm}}

\begin{document}
\begin{frontmatter}
\title{The ATLAS pixel detector}
\author{Markus Cristinziani\thanksref{Someone}}
\ead{Markus.Cristinziani@cern.ch}
\address{Physikalisches Institut, Nussallee 12, 53115 Bonn, Germany}
\thanks[Someone]{for the ATLAS pixel collaboration}

\begin{abstract}
After a ten years planning and construction phase, the ATLAS
pixel detector is nearing its completion and is scheduled to be integrated
into the ATLAS detector to take data with the first LHC collisions in 2007.
An overview of the construction is presented with particular emphasis on some
of the major and most recent problems encountered and solved.
\end{abstract}
\end{frontmatter}

\section{Introduction}
The LHC proton-proton collider is expected to operate at a center-of-mass
energy of \vu{14}{\TeV}, a bunch-crossing rate of \vu{40}{MHz} and a design luminosity of
\vu{10^{34}}{cm^{-2}\,s^{-1}}. With the data recorded by the multi-purpose detectors,
ATLAS and CMS, the mechanism of electroweak symmetry breaking and physics beyond 
the Standard Model will be explored.
The high radiation environment and the large data rate pose severe constraints
on the detector technology, in particular for the inner detectors. In ATLAS,
the inner detector consists of a pixel inner tracking subsystem, surrounded by
a silicon microstrip and a transition radiation trackers.
The lifetime equivalent
neutron dose which need to be sustained by the pixel detector is \vu{50}{Mrad} or $10^{15}$
neutron equivalent.

\section{Overview}
The pixel detector~\cite{TDR98} is arranged in a cylindrical symmetry around 
the beam pipe (barrel)
and in addition two end-cap subsystems (plugs) in the forward and backward region. 
The three barrel layers are located  at a distance of \vu{5}{cm}, \vu{9}{cm} and \vu{12}{cm} from the beam axis and
are equipped with 1456 partially overlapping identical pixel modules. Each endcap 
consists of three parallel planes at a nominal distance of \vu{50}{cm}, \vu{58}{cm} 
and \vu{65}{cm} from
the interaction point and houses 288 modules. This geometrical arrangement allows 
a coverage in pseudorapidity for tracks with $|\eta| < 2.5$.

\section{Modules}

\begin{figure}[htbp] \vups
\begin{center}
\includegraphics[width=0.88\columnwidth]{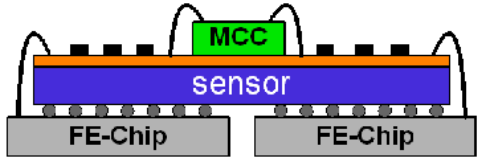}
\end{center}
\caption{Schematic cross section of a pixel module unit.}
\label{f1} \volp
\end{figure}

The basic unit of the pixel detector is a module. It consists of a silicon sensor,
16 front-end read-out chips arranged in two rows of 8 chips, a Kapton flex circuit
with the Module Controller Chip and a pigtail connector. A schematic cross section 
view of a pixel module is shown in Fig.~\ref{f1}.

The sensor~\cite{sensor} has an active area of \vu{60.8}{mm} $\times$ \vu{16.4}{mm}.
The 47268 pixels are implemented as $n^+$ implants on the read-out side
in \vu{250}{\mu m} thick oxygenated float-zone silicon n-bulk material. 
Radiation damage will type invert
the sensor bulk and then increase the depletion voltage. A multiple 
guard-ring structure on the back side of the sensor allows for a maximum
bias voltage of \vu{600}{V}. This will provide nearly full depletion even after 
ten years operation in the LHC environment.

\begin{figure}[htbp] \vups
\begin{center}
\includegraphics[width=0.88\columnwidth]{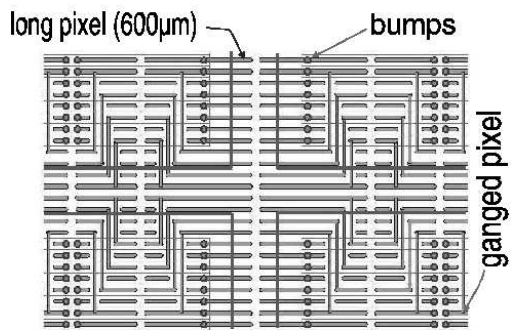}
\end{center}
\caption{Pixel cell design on the silicon sensor. The standard design in modified
in the inter-chip region (long pixels and ganged pixels) in order to maximize acceptance.}
\label{f2} \volp
\end{figure}

Each pixel cell has the dimensions \vu{50}{\mu m} $\times$ \vu{400}{\mu m} which will provide 
a point resolution of \vu{10}{\mu m} in the $r\phi-$coordinate~\cite{testbeam}. 
In the regions between front-end chips, pixels have either 
a modified geometry (\vu{50}{\mu m} $\times$ \vu{600}{\mu m}) or are connected with each other,  
such that, at the cost of ambiguous reconstruction of hit positions, 
there is no dead area on the sensor surface (Fig.~\ref{f2}). 

The silicon sensor is connected to the read-out front-end chips through fine pitch
bump bonding with the flip-chip technique to form a bare module. The bump bonds
provide electrical,
mechanical and thermal contact at the same time. This fabrication step was done by two
different providers, IZM and AMS, with PbSn and In technology, respectively.

The front-end chip FE-I3 is described in detail elsewhere~\cite{Ivan}. It is implemented
in a standard \vu{0.25}{\mu m} CMOS process with a radiation tolerant layout, which has been
demonstrated up to \vu{100}{Mrad} of total dose. It contains 2880 read-out cells, arranged
in a $18\times 160$ matrix matching the sensor pixel geometry. In the analog section the 
charge deposited in the sensor is amplified and compared to an individually tunable threshold
by a discriminator. The digital readout buffers the pixel address, a time stamp, and 
the signal amplitude as time-over-threshold (ToT) of hits. Hits which are selected by
trigger signals within the Level 1 latency (\vu{3.2}{\mu m}) are read-out, 
otherwise they are deleted.

\begin{figure}[htbp] \vups
\begin{center}
\includegraphics[width=0.88\columnwidth]{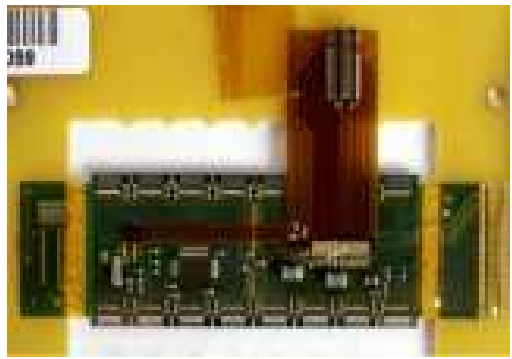}
\end{center}
\caption{A completely assembled module inside a flex holder used for safe manipulation during tests and transportation.}
\label{f3} \volp
\end{figure}

The last step of module assembly (``dressed module'') consists in gluing a flexible 
Kapton (flex) printed-circuit board
to the back side of the bare module
and connecting it through ultrasonic micro-wirebonds (Fig.~\ref{f3}). 
The flex contains passive components and the Module Controller Chip (MCC)~\cite{MCC} which steers the
communication between the data acquisition system and the front-end chips. Event building and 
error handling are managed at this stage by the MCC. 

All module components have been extensively tested in irradiation runs to ensure 
that the operation of the pixel detector will be possible even after the expected lifetime
dose of \vu{50}{Mrad}. Seven fully assembled production modules were irradiated and tested
in the laboratory.
A typical noise increase of only 10\% was observed. Other modules were characterized in a 
test beam, showing an almost fully depleted sensor, slightly reduced charge collection efficiency
due to trapping, and excellent performance in high rate tests. See ref.~\cite{testbeam} for details.

\section{Production and integration}

Approximately $70,\!000$ front-end chips have been produced on 250 wafers with a 
yield exceeding $80\%$.
Extensive testing was performed on each wafer before and after bumping, thinning down to \vu{190}{\mu m}
and dicing. 
At the bare module level, the front-end chips were again tested, in particular to detect 
defective bump bonds. A reworking procedure has been developed to recover these modules
with high efficiency. Approximately 10\% needed to undergo reworking. The recovery efficiency
was 90\%, averaged over the two vendors.

Approximately 2000 modules have been assembled and tested at five production sites (Fig.~\ref{f3}).
Each module receives a ranking penalty measured in equivalent dead channels (edc) which determines the
later placement of the module in the detector. Only 4\% of the modules were rejected at this stage.

Laboratory measurements on production modules are described in detail elsewhere~\cite{Jorn}.
For the threshold scan different charges are artificially injected and the number of hits recorded.
Thus the threshold of the discriminator can be determined and adjusted for each individual channel.  
The timewalk is measured to ensure that hits can be associated to the correct bunch
crossing during normal data taking at LHC. The module is then illuminated with a radioactive source
to determine dead or inefficient pixels.
The ToT response is calibrated into charge using the on-chip injection circuit and verified to
agree to expectations based on the radioactive source measurement.

\begin{figure}[htbp] \vups
\begin{center}
\includegraphics[width=0.88\columnwidth]{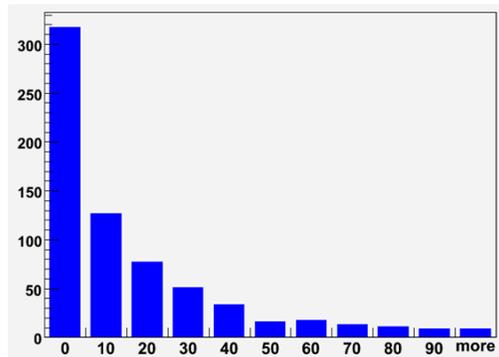}
\end{center}
\caption{Number of modules vs. dead channels for high-quality modules available for the
innermost barrel layer.}
\label{f4} \volp
\end{figure}

Figure~\ref{f4} shows a distribution of dead channels for the modules chosen for the innermost layer.
More than 300 modules resulted with less than 10 dead channels.

\begin{figure}[htbp] \vups
\begin{center}
\includegraphics[width=0.88\columnwidth]{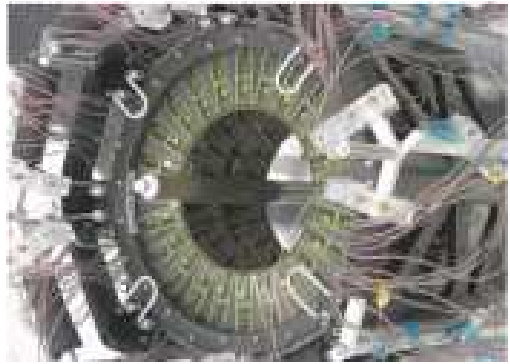}
\end{center}
\caption{One pixel endcap.}
\label{f5} \volp
\end{figure}

\begin{figure}[htbp] \vups
\begin{center}
\includegraphics[width=0.88\columnwidth]{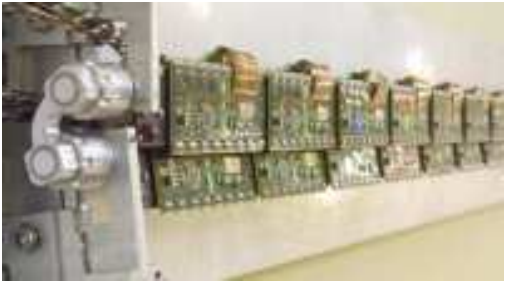}
\end{center}
\caption{View of a bistave, two barrel staves connected by the same cooling line.}
\label{f6} \volp
\end{figure}

The modules are loaded onto carbon-fiber structures in three sites. For the endcap (Fig.~\ref{f5}),
six modules are mounted on a sector assembly plate corresponding to 1/8 of a disk. Two subsequent disks
are rotated to optimize coverage. 
For the barrel (Fig.~\ref{f6}), 13 modules are precisely glued on a stave.
A quick connectivity test for all modules is performed to exclude damage during loading. The 
components are then sent to CERN for final integration.

\begin{table}[htbp] \vups
\begin{center}
\begin{tabular}{l@{\hspace{1ex}}|@{\hspace{1ex}}c@{\hspace{1ex}}|@{\hspace{1ex}}c}
layer & barrel & endcaps\\
\hline
inner & 0.13\% & 0.04\% \\
center & 0.17\% & 0.16\% \\
outer & 0.27\% & 0.22\% \\
\end{tabular}
\end{center}
\caption{Fraction of dead channels per module, averaged over all modules integrated into the detector.}
\volp
\end{table}

A maximum ranking value of 60 edc was required for a module 
to be integrated on a stave for the innermost layer.
Table 1 shows the fraction of dead channels per module, 
averaged over all modules which have been integrated into the
detector.

\section{Integration status}

\begin{figure}[htbp] \vups
\begin{center}
\includegraphics[width=0.88\columnwidth]{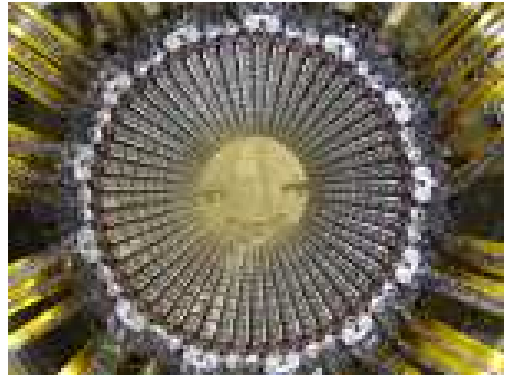}
\end{center}
\caption{View of the completely assembled outermost layer of the barrel. The inner layers have also been assembled and inserted into this one.}
\label{f7} \volp
\end{figure}

At the time of writing the three barrel layers and the two endcaps
have been fully assembled. One endcap is being tested
on a cosmic-ray setup, while the barrel layers are arranged in 
their final position around the beam pipe. Fig.~\ref{f7} shows the 
completed outermost barrel layer. All modules are connected
and tested to work properly with minor degradation with respect to
the previous laboratory tests. In parallel, a system test has been 
setup to assess the read-out and DAQ readiness~\cite{system}.

During construction, a number of unforeseen issues and potential
bottlenecks for the final detector completion arose and solutions
were found.

To protect the delicate wirebonds of the front-end chips, of the pigtail
connection and of the MCC, a protective coating (potting) is applied to 
prevent accidental mechanical breakage. During thermal stress tests
a number of bonds connecting the MCC to the flex broke. While thorough
studies demonstrated that a similar effect is not observed for the 
other potted bonds, it was decided to manually remove the potting on 
the MCCs and rebond them, which was achieved with 100\% efficiency, even
on dressed modules.

\begin{figure}[htbp] \vups
\includegraphics[width=0.66\columnwidth]{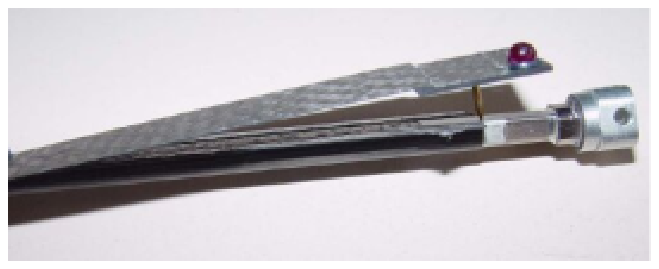}
\includegraphics[width=0.33\columnwidth]{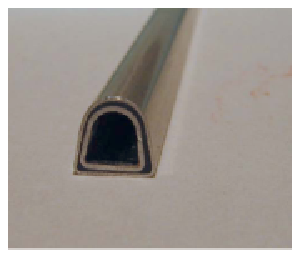}
\caption{Example of a delaminated stave (left). The pipe insertion solution
to the leak problem (right).}
\label{f8} \volp
\end{figure}

Two problems connected to the stave design were detected at a relatively
late stage of production. Owing to the non-circular cross-section of the
cooling pipe and the pressure of the coolant and the particular linear
geometry of the stave, delamination of the carbon-carbon structure from 
the cooling pipe has been observed. This problem was resolved by adding a 
peek collar to the stave extremities (Fig.~\ref{f8} left).
Secondly, some cooling pipes were discovered to be leaky due to corrosion.
The cause was found to be the brazing of the cooling fitting followed 
by a not accurate enough quality control (water vapor in the pipe). 
Cooling tubes and fittings were designed and laser welded in a new
production. For existing, unloaded staves, the cooling pipe was replaced.
For the approximately 40 already loaded staves, new pipes were inserted 
into the old ones (Fig.~\ref{f8} right) providing a still satisfactory thermal 
contact.

\begin{figure}[htbp] \vups
\includegraphics[width=0.49\columnwidth]{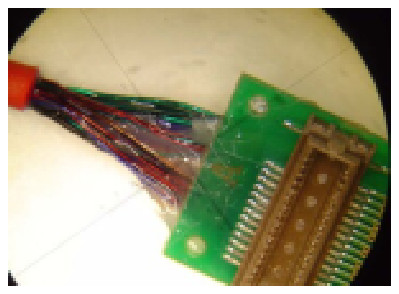}
\includegraphics[width=0.49\columnwidth]{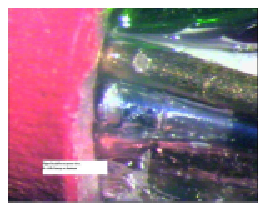}
\caption{Low mass cable in the region of the connector (left). Cracks in the
insulation are evident by closer inspection (right).}
\label{f9} \volp
\end{figure}

During loading of the first barrel staves, failures of some 
low-mass cables were observed at the pigtail connector junction due to excessive 
stress on the \vu{100}{\mu m}-\vu{300}{\mu m} thick wires. After closer 
inspection, cracks in the insulation of the power cables (Fig.~\ref{f9})
were discovered in an important fraction of the cables. 
The specific production technique was identified as the origin of the failure, and 
was corrected in time. A second batch of cables produced with the rectified
process did not exhibit this problem.

\section{Summary}
A short description of the ATLAS pixel detector has been presented. 
The radiation environment and occupancy demanded the development of new technologies
that started more than ten years ago. The project recently finished the 
module production and integration phase. Several issues were discovered
and solved. 
Installation of the three-layer pixel detector is currently planned by April 2007
and is on schedule.

The three-layer pixel detector is currently expected to be 
installed as the last sub-detector into ATLAS on schedule by April 2007.

\section{Acknowledgements}
The development and construction of the \mbox{ATLAS} pixel detector involves more than
100 dedicated scientists and engineers from Berkeley, Bonn, Dortmund, Geneva, Genoa, 
Marseilles, Milan, New Mexico, Ohio, Oklahoma, Prague, Siegen, Udine and Wuppertal.
This work is partially funded by BMBF under contract 05 HA4PD1/5.

\end{document}